# Threshold Voltage Improvement and Leakage Reduction of AlGaN/GaN HEMTs Using Dual-Layer SiN$_x$ Stressors

Wei-Chih Cheng, *Student member, IEEE*, Minghao He, Siqi Lei, Liang Wang, Jingyi Wu, Fanming Zeng, Qiaoyu Hu, Feng Zhao, *Senior member, IEEE*, Mansun Chan, *Fellow, IEEE*, Guangrui (Maggie) Xia, and Hongyu Yu, *Senior member, IEEE*

*Abstract*—In this work, AlGaN/GaN HEMTs with dual-layer SiN$_x$ stressors (composed of a low-stress layer and a high-stress layer) were investigated. The low-stress padding layer solved the surface damage problem caused during the deposition of the high-stress SiN$_x$, and provided a good passivated interface. The HEMTs with dual-layer stressors showed a 1 V increase in the threshold voltage ($V_{th}$) with comparable on-current ($I_{on}$) and RF current gain to those without stressors. Moreover, the off-current ($I_{off}$) was shown to be reduced by one to three orders of magnitude in the strained devices as a result of the lower electric field in AlGaN, which suppressed the gate injection current. The dual-layer stressor scheme supports strain engineering as an effective approach in the pursuit of the normally-off operation of AlGaN/GaN HEMTs.

## I. Introduction

GaN-based high electron mobility transistors (HEMTs) have played an increasingly important role in power electronics and high frequency applications [1], [2]. A high drive current can be achieved due to the high saturation velocity of electrons in GaN and the presence of two-dimensional electron gas (2DEG) at the AlGaN/GaN heterojunction [3]. Normally-off operation is preferred for radio-frequency (RF) power amplifiers to simplify the design of the whole RF system. However, major normally-off device options, such as gate-recessed metal-insulator-semiconductor HEMT (MIS-HEMT) [4] or HEMT with p-type GaN gate [4], [5], require etching processes, which introduce surface damages that may deteriorate the amplifying performance of the devices [6]. Even though the cyclic oxidation/wet-etching digital etching processes can effectively avoid the etching caused damage and realize high-performance devices [], [], this technique is hard to be used for industrial application.

Due to the piezoelectric nature of GaN, the 2DEG concentration and thus the electrical characteristics of the AlGaN/GaN HEMT can be modulated by applying stress [9], [10]. Besides, experimental work and simulation results showed that the stress provided by the dielectrics liner can induce significant amount of piezoelectric charges in the AlGaN/GaN heterostructure underneath the submicron or smaller gate metal [9], [11]. Therefore, strain engineering is believed to be an effective approach to adjust the threshold voltage ($V_{th}$) of an AlGaN/GaN HEMT with a scaled gate length.

Silicon nitride (SiN$_x$) liners deposited using plasma enhanced chemical vapor deposition (PECVD) with dual plasma excitation frequencies have been reported as effective stressors [12]. Increasing the duty cycles of low frequency plasma excitation is useful to deposit highly compressively-stressed SiN$_x$, but it also introduces more nitrogen ion bombardment onto the semiconductor surface, resulting in more surface damages [11], [13]. In previous work, the $V_{th}$ of AlGaN/GaN HEMTs with gate length $L_g$ = 0.1 µm was increased by over 1 V using compressive SiN$_x$ liners [11]. However, the semiconductor surface damages caused during the deposition of compressive SiN$_x$ degraded the electrical performance of devices, such as the on-current ($I_{on}$) decrease and off-current ($I_{off}$) increase [11].

This work was supported by Grant #2019B010128001 from Guangdong Science and Technology Department, Grant #JCYJ20160226192639004 and # JCYJ20170412153356899 from Shenzhen Municipal Council of Science and Innovation. Corresponding authors: Guangrui (Maggie) Xia and Hongyu Yu.

Wei-Chih Cheng is with the School of Microelectronics, Southern University of Science and Technology (SUSTech), Shenzhen 518055, Guangdong, China, and also with the Department of Electronic and Computer Engineering, Hong Kong University of Science and Technology (HKUST), Hong Kong (e-mail: weichih.cheng@connect.ust.hk).

Minghao He, Siqi Lei, Liang Wang, Jingyi Wu, Fanming Zeng, and Qiaoyu Hu are with the School of Microelectronics, Southern University of Science and Technology (SUSTech) , Shenzhen 518055, China.

Feng Zhao is with the School of Engineering and Computer Science, Washington State University, Vancouver, WA 98686, USA (e-mail: feng.zhao@wsu.edu).

Mansun Chan is with the Department of Electronic and Computer Engineering, Hong Kong University of Science and Technology (HKUST), Hong Kong.

Guangrui (Maggie) Xia is with the School of Microelectronics, Southern University of Science and Technology (SUSTech), Shenzhen 518055, China, and also with the Department of Materials Engineering, The University of British Columbia, Vancouver, BC, Canada (e-mail: gxia@mail.ubc.ca).

Hongyu Yu is with the School of Microelectronics, Southern University of Science and Technology (SUSTech), Shenzhen 518055, Guangdong, China, also with GaN Device Engineering Technology Research Center of Guangdong, Shenzhen 518055, Guangdong, China, and also with The Key Laboratory of the Third Generation Semi-conductor, Shenzhen 518055, Guangdong, China (e-mail: yuhy@sustech.edu.cn).

To overcome the surface damage problem with the single-layer compressive $SiN_x$, in this work, dual-layer $SiN_x$ stressors were investigated. A dual-layer $SiN_x$ stressor was composed of a low-stress $SiN_x$ interlayer for interface passivation and a high-stress $SiN_x$ layer for compressive stress. The AlGaN/GaN RF HEMTs with 0.1 μm gate length for RF application were prepared, and their $V_{th}$ were increased by the dual-layer $SiN_x$ stressors. The dual-layer scheme was shown to effectively prevent the semiconductor surface from the nitrogen ion bombardment during the deposition of compressive $SiN_x$. Besides the increased $V_{th}$, the strained devices with dual-layer stressors also showed $I_{on}$ and high-frequency performance comparable to those of the baseline devices with optimal passivation scheme. Moreover, the $I_{off}$ in the strained devices with the dual-layer stressors was lower than in the baseline devices. The $V_{th}$ increase without etching process and the compromise of electrical performance presents an alternative to pursue normally-off RF AlGaN/GaN HEMT.

## II. $V_{th}$ OF GaN BASED HEMT

The threshold voltage ($V_{th}$) of AlGaN/GaN HEMTs can be expressed by a polarization dependent model, which is given as [14]

$$V_{th}(x) = \varphi_m(x) - \Delta E_c(x) - \frac{qN_Dd}{2\epsilon(x)} - \frac{\sigma(x)}{\epsilon(x)}d \quad (1)$$

where $x$ is the aluminum mole fraction of AlGaN, $\varphi_m(x)$ is the barrier height of the Schottky gate, and $\Delta E_c(x)$ is conduction band discontinuity between $Al_xGa_{1-x}N$ and GaN. $N_D$, $d$, and $\epsilon(x)$ are doping concentration, thickness, and dielectric permittivity of the $Al_xGa_{1-x}N$ barrier, respectively. $\sigma(x)$ is the total polarization induced sheet charge density, which is given as

$$\sigma_x = P_{sp}(Al_xGa_{1-x}N) - P_{sp}(GaN) + P_{pz}(Al_xGa_{1-x}N) - P_{pz}(GaN) \quad (2)$$

where $P_{sp}$ and $P_{pz}$ are the spontaneous polarization and the piezoelectric polarization in the AlGaN and GaN layers, respectively. Equation (1) and (2) ignored the effects from the GaN cap and AlN spacer for simplification. In the AlGaN/GaN heterostructures, the piezoelectric polarization of AlGaN is caused by the tensile strain due to the lattice constant mismatch between AlGaN and GaN, and the GaN layer is considered to be fully relaxed due to the thick buffer layers. By introducing in-plane compressive stress, the in-plane tensile strain in AlGaN can be neutralized, and the total polarization reduces, resulting in an increase in the $V_{th}$ of AlGaN/GaN HEMT.

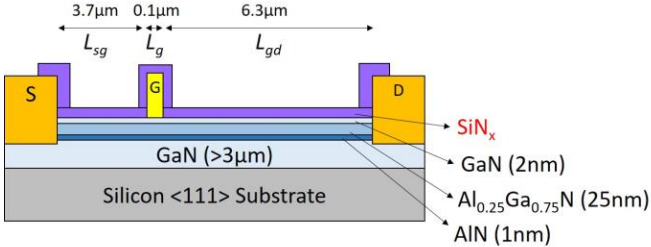

Fig. 2 Device structure of AlGaN/GaN HEMT with gate length ($L_g$), source-to-gate length ($L_{sg}$), and gate-to-drain length ($L_{gd}$) labelled.

TABLE I
DEVICE MATRIX IN THIS WORK

| Device group | Low-stress (0.3 GPa) $SiN_x$ for surface Passivation (nm) | High-stress (-1 GPa) $SiN_x$ for stress introduction (nm) |
|---|---|---|
| 1. Baseline devices | 200 | - |
| 2. Strained devices | - | 200 |
| | 3.5 | 200 |
| | 7 | 200 |
| | 10.5 | 200 |
| | 14 | 200 |

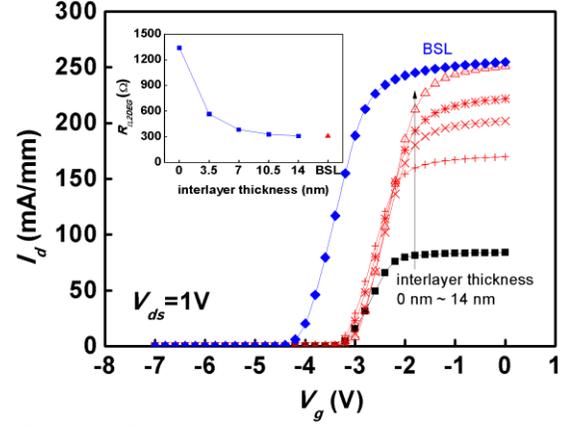

Fig. 1 Transfer characteristics of the strained devices with different interlayer thicknesses (0, 3.5, 7, 10.5, 14 nm) in comparison with a baseline device (BSL). The strained devices showed 1 V higher $V_{th}$ than the baseline device. The device with thicker interlayer shows higher on-current ($I_{on}$), consistent with the trend of 2DEG sheet resistance ($R_{\square,2DEG}$) extracted using transmission line model (inset).

[13]

## III. AlGaN/GaN HEMT WITH $SiN_x$ STRESSORS

The structure of the AlGaN/GaN HEMTs is shown in Fig. 1. The 6 inch Si wafer with MOCVD grown GaN/AlGaN/AlN/GaN epitaxy was purchased from a commercial supplier. The epitaxial structure consists of a 1.05 μm high-resistivity GaN buffer, a 700 nm i-GaN channel, a 0.8 nm AlN spacer, a 19 nm $Al_{0.25}Ga_{0.75}N$ barrier, and a 3 nm GaN cap. The device fabrication started with the device isolation using a $BCL_3/Cl_2$ based inductively coupled plasma dry etching. Then, Ti/Al/Ti/Au-based metal stack was deposited using e-beam evaporator and annealed at 830°C in nitrogen ambient to form the source/drain Ohmic contacts. After preparing the Ni/Au-based metal gate using e-beam lithography and e-beam evaporator, the $SiN_x$ layers were deposited by PECVD with dual plasma excitation frequencies. Subsequently, metal pads was deposited after via opening.

In this work, two $SiN_x$ deposition conditions were developed. In the baseline devices, the low-stress-$SiN_x$ layer was optimized for passivation to improve the drive current, which had an unintentional stress of 0.3 GPa [11]. This deposition condition was also used to deposit the low-stress interlayer for surface passivation in the dual-layer stressor scheme. For the high-stress layer, an increased duty cycles of low-frequency plasma excitation was used to introduce -1 GPa compressive stress to increase the $V_{th}$ of devices. The stress values were obtained from the curvature measurements on blanket $SiN_x$ films on Si substrates. In this work, two groups of HEMTs were fabricated: the baseline devices had a low-stress $SiN_x$ layer per device, and the second group (strained devices) had a high-stress layer and may have a low-stress layer as an interlayer per device. In the second group with dual-layer stressors, different low-stress layer thicknesses were designed as listed in Table 1, and the incorporation of the low-stress layer was proved not influence the overall intrinsic stress of the dual-layer stressors. In this work, Group 1 and Group 2 devices will be called baseline devices and strained devices, respectively, even though the $SiN_x$ layer applied onto Group 1 devices were also slightly stressed (0.3 GPa). The thicknesses were calculated by the deposition rate and time.

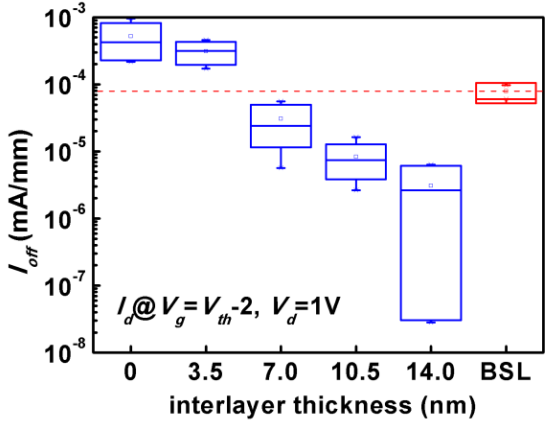

Fig. 4 Statistics of off-current ($I_{off}$) of the strained devices with different interlayer thicknesses and the baseline devices. The strained devices with thicker interlayers showed lower $I_{off}$. The strained devices with 7 nm interlayer or the thicker ones showed lower $I_{off}$ than the baseline devices.

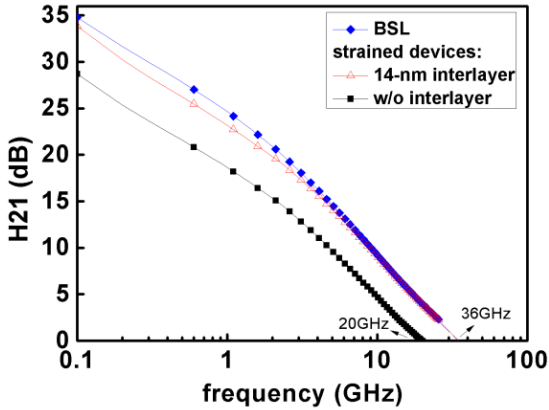

Fig. 3 H21 current gain of a typical baseline device (BSL) and strained devices with and without interlayers. The incorporation of interlayer improved RF performance of strained devices, and the strained devices with 14 nm interlayer showed comparable performance as the baseline device, attaining the cutoff frequencies ($f_t$) of 36 GHz. The devices were tested at $V_d$ = 7V and $V_g = V_{th} + 1$ V for each device.

## IV. ELECTRICAL CHARACTERISTICS

The transfer characteristics of strained devices were shown in Fig. 2, and the curve of baseline devices was also shown for comparison. Compared with the baseline devices, the $V_{th}$ of strained devices were successfully increased by 1 V, because the compression depleted the 2DEG underneath the gate region, consistent with the previous results [11]. The $I_{off}$ was extracted, as shown in Fig. 3, which was defined as the drain current at gate bias $V_g = V_{th}$-2 V and drain voltage $V_d$ = 1 V. Consistent with the previous study in [11], strained devices without interlayer showed reduced $I_{on}$ and higher $I_{off}$ because of the damaged semiconductor surface from the nitrogen ion bombardment during the deposition of the high-stress $SiN_x$ [11], [13]. For strained devices with interlayers, thicker interlayers effectively increased $I_{on}$ and decreased $I_{off}$, indicating that the interlayer effectively protected the semiconductor surface during the deposition of the high-stress $SiN_x$. The sheet resistance of 2DEG ($R_{\square,2DEG}$) extracted using the transmission line model was also shown in the inset of Fig. 2, showing a consistent trend as $I_{on}$ of devices. The strained device with 14 nm interlayer showed a similar $I_{on}$ and a lower $I_{off}$ compared with the baseline devices, suggesting the reduction of surface damages. The reduction of semiconductor damages also translated to the better RF performance of strained devices. Fig. 4 plots the current gains (H21) of devices at $V_d$ = 7 V and $V_g = V_{th}$+1 V. Strained devices with 14 nm interlayers showed a similar H21 current gain to the baseline devices, with the cutoff frequency ($f_t$) of 36 GHz, while that without the interlayer showed lower h21 gain and $f_t$ of 20 GHz.

It was shown that lower $I_{off}$ was attained with thicker interlayer. The higher $I_{off}$ of strained devices without the interlayer or with the thin interlayers were believed to have been caused by the surface damages during the high-stress $SiN_x$ deposition. However, strained devices with 7 nm or thicker interlayer showed even lower $I_{off}$ compared with the baseline devices, indicating that the passivation quality was not the only factor determining $I_{off}$ in this study.

From the above data, the devices with 14 nm interlayers had the best performance (similar $I_{on}$ and RF performance, one to three orders of magnitude lower $I_{off}$ and 1 V higher $V_{th}$ compared with the baseline devices). Therefore, these devices were further investigated and discussed in the following sections for the stress effect on AlGaN/GaN HEMTs. For simplification, strained devices with specifically 14 nm interlayers will be called strained devices in the following sections.

## V. STRESS AND DEVICE MODELLING AND SIMULATIONS

To further investigate the influence of the applied stress on 2DEG concentration, the distribution of the external stress and the electron concentration were simulated using two finite element analysis tools, Victory Stress and Victory Device, developed by Silvaco,.

Fig. 5(a) and (b) showed the stress simulation results with the measured $SiN_x$ intrinsic stress values of 0.3 GPa for the low-stress layer and -1 GPa for the high-stress layer, corresponding to the baseline devices and the strained devices. The high-stress layer introduced an in-plane compressive stress (up to -1.2 GPa) underneath the gate area and an negligible tensile stress (< 50 MPa) in the access area, as shown in Fig. 5(b).

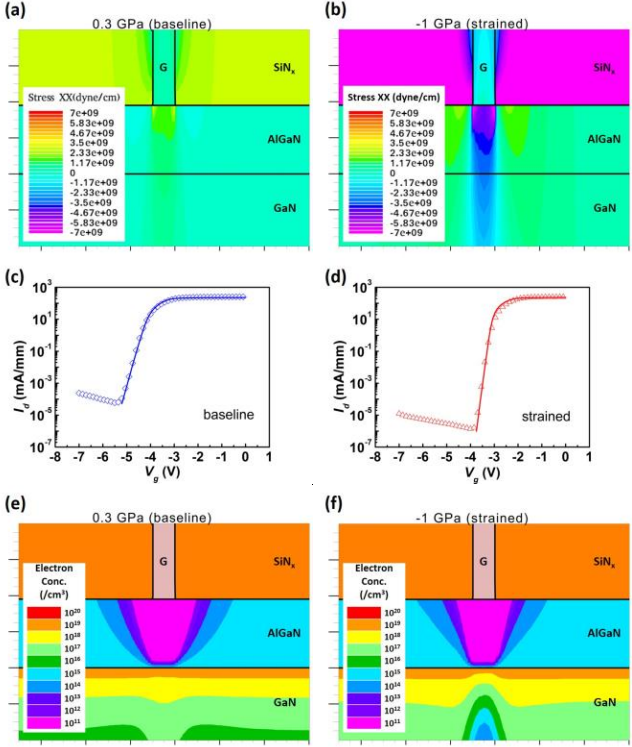

Fig. 6 Simulation and fitting results with Victory Stress and Victory Device. (a)(b) Simulated stress distribution, (c)(d) subthreshold characteristics (solid lines: best fitting curves and dots: experimental measurements), and (e)(f) electron concentration of the device with a low-stress (0.3 GPa) layer and a high-stress (-1 GPa) layer, respectively. The high-stress layer caused compression and electron depletion underneath gate. The device with a high-stress layer showed a positive $V_{th}$ shift compared with that with a low-stress layer only (the baseline device).

The device model was calibrated to match the subthreshold characteristics of the baseline devices, as shown in Fig. 5(c). The trap densities and the relative ratio between the applied stress induced, lattice mismatch induced, and spontaneous piezoelectric effects were used as the fitting parameters. The best fitting parameters generated good fits to the experimental subthreshold I-V data, showing a 1 V $V_{th}$ shift when the -1 GPa high-stress layer was incorporated, as shown in Fig. 5(d). The $V_{th}$ shift was a result of the electron depletion caused by the applied compressive stress. Due to the piezoelectric nature of AlGaN, the compressive stress neutralize the original piezoelectric polarization in AlGaN, which is $P_{pz}$(AlGaN) in Section II, reducing the concentration of 2DEG, as shown in Fig. 5(f). In contrast, the low-stress layer only caused a slight electron accumulation underneath the gate, as shown in Fig, 5(e).

## VI. REDUCED $I_{off}$ OF STRAINED DEVICES

The gate length of the aforementioned devices were in submicron scale ($L_g = 0.1$ μm), and the electrical characteristics of submicron or even smaller device can be effectively changed by stress layers according to the edge force model [9]. To verify that the reduced $I_{off}$ of the strained devices was resulted from the applied stress, we prepared the devices with $L_g$ = 2 micron-scale gate length (micron devices). Due to the much larger sizes, the amount of stresses that can be introduced was much less than that of submicron devices, and thus the electrical characteristics of micron devices were proved to be influenced less by the applied stress [9], [11]. The structure of the micron devices was similar to that shown in Figure. 1, with $L_g$ of 2 μm and $L_{gd}$ of 10 μm. The transfer characteristics of the submicron devices and micron devices at $V_d = 1$V were shown in Fig. 6(a) and (b), respectively. Unlike the submicron devices, the micron devices showed no obvious $I_{off}$ difference with and without the stressors. The reverse Schottky gate leakage of the devices showed the similar results, as shown in Fig. 6(c) and (d). The different effects of stressors on $I_{off}$ between micron devices and submicron devices suggested that the lower $I_{off}$ of the strained submicron device was caused by the applied stress. Besides, the reverse Schottky gate current $I_g$ of the submicron devices overlapped when bias was below –4 V. It implied potentially the same leakage mechanism dominating the leakage of two devices under the low reverse bias condition, and another mechanism of leakage took place in the baseline devices when higher reverse bias was applied.

In Fig. 6(c) and (d), except the submicron strained device, all devices showed strong bias-dependent leakage current in certain bias regions, as indicated with each curve. The strong bias-dependent leakage current of GaN based HEMT could be dominated by Fowler–Nordheim (FN) tunneling [15]. The dependence of FN tunneling current density ($J_{FN}$) on the barrier electric field ($E$) is given by

$$J_{FN} = AE^2 exp(-B/E) \qquad (3)$$

with

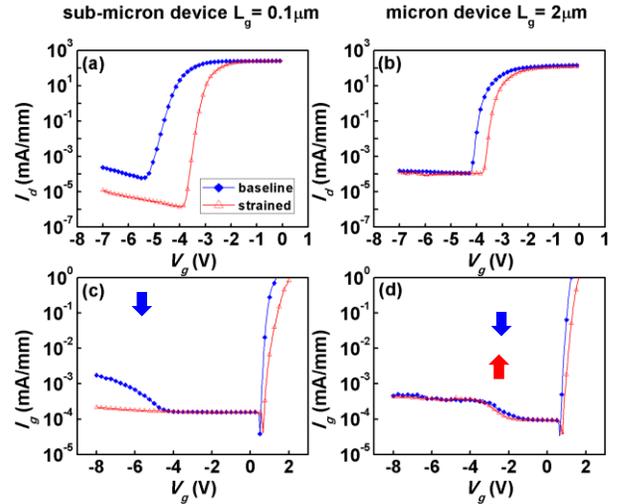

Fig. 5 (a)(b) Transfer characteristics of the devices at $V_d = 1$ V with different gate lengths and different strain conditions. The submicron devices showed different level of $I_{off}$ when stress was introduced while the micron devices showed the same level. (c)(d) Reverse Schottky gate leakage showed the similar phenomena as $I_{off}$ in the transfer curves. The submicron baseline device and two micron devices showed bias-dependent Schottky leakage current in indicated bias regions.

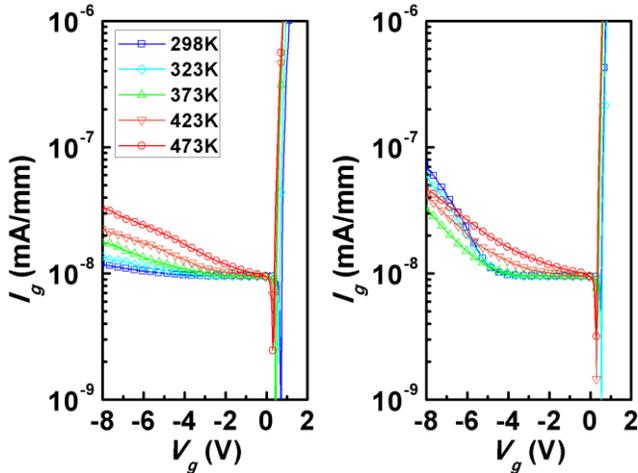

Fig. 7 Temperature dependence of reverse Schottky gate leakage of a baseline (left) and a strained (right) device. Reverse current of the baseline device showed weaker dependence on the temperature compared with the strained device.

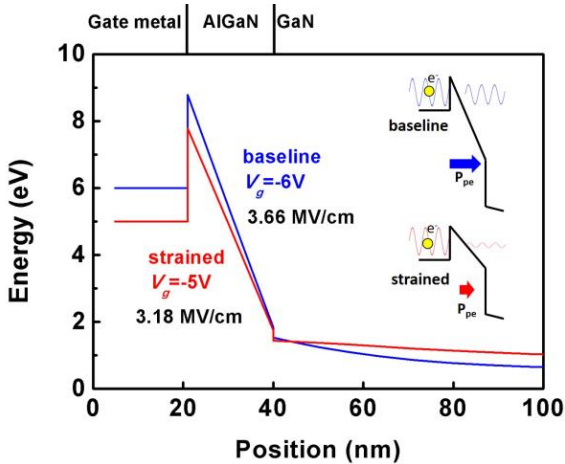

Fig. 8 Conduction band diagram of gate metal/AlGaN/GaN system with 0.3 GPa (baseline) and -1 GPa (strained) stressor at $V_g = V_{th}$-2V. With less piezoelectric polarization, the effective barrier thickness is thicker for the strained device, and possibility of FN tunneling is less (inset).

$$B = 8\pi\sqrt{2m_n^*(q\varphi_{eff})^3}/3qh \quad (4)$$

where $A$ is a constant, $E$ is the internal electric field of the barrier (AlGaN in this system), $m_n^*$ the effective mass of the semiconductor, $h$ the Planck's constant, and $\varphi_{eff}$ is the effective barrier height. FN tunneling current is independent of temperature according to (3). As shown in Fig. 7, the leakage current of the submicron baseline device did not change obviously and showed no strong correlation with temperature in the strong bias-dependent leakage region, suggesting the dominance of FN tunneling. In contrast, the reverse current of the submicron strained device increased with temperature significantly, indicating the absence of FN tunneling occurrence. To investigate the reason why FN tunneling was suppressed by the stress, the conduction band diagram was simulated using Silvaco Victory Device with the model parameters calibrated with the subthreshold characteristics discussed in section V.

The conduction band diagrams in the gate region of the baseline device and the strained device at $V_g = V_{th}$-2 V were simulated, as shown in Fig. 8. The internal electrical field strengths in AlGaN were 3.66 MV/cm and 3.18 MV/cm for the baseline device and strained device, respectively. For a strained device, the compression neutralized the inherent piezo polarization caused by the lattice mismatch at the heterojunction. As a result, the internal electrical field in AlGaN was reduced, and FN tunneling was hence suppressed, reducing the gate leakage of the strained devices.

VII. CONCLUSION

Due to the piezoelectric nature of GaN, a 1 V increase in $V_{th}$ of AlGaN/GaN HEMTs was achieved by introducing compressive stresses with the dual-layer $SiN_x$ stressors. The low-stress $SiN_x$ layer in the dual-layer $SiN_x$ stressor scheme successfully retained the similar $I_{on}$ as the baseline devices by avoiding the surface damages caused during the deposition of the highly compressive $SiN_x$ layer. As a result, the comparable passivation quality translated to the comparable high-frequency performance of the strained devices and baseline devices. Moreover, the applied compression reduced the internal electric field in AlGaN, suppressing the leakage current $I_{off}$ caused by FN tunneling when a strained device is under a reverse bias. This $V_{th}$ increase without a recess etching process or any compromises of the electrical performance, such as $I_{on}$, $I_{off}$, and RF performance, supports strain engineering as an effective approach in pursuing enhancement-mode AlGaN/GaN HEMTs for RF applications.

ACKNOWLEDGMENTS

The work was conducted at Materials Characterization and Preparation Center (MCPC) at SUSTech, and we acknowledge the technical support from the staff and engineers at MCPC. Wei-Chih Cheng would also like to thank his labmates Marzieh Mahrokh and Robert Sokolovskij for practical suggestions on this article.